\documentclass[aps,prl,twocolumn,superscriptaddress,showpacs,floatfix]{revtex4-2}

\usepackage{amsmath,amssymb,amsfonts,float,graphics,epsfig,epstopdf,color,verbatim,tabularx,bm,multirow,appendix}
\usepackage[utf8]{inputenc}
\usepackage[T1]{fontenc}
\usepackage{xcolor}
\usepackage{dsfont}
\usepackage{textcomp}
\usepackage{yfonts}
\usepackage{footnote}
\usepackage{bm}
\usepackage{subfigure}
\usepackage{mathrsfs}
\usepackage{graphicx}
\usepackage{verbatim}
\usepackage[colorlinks=true, citecolor=blue, linkcolor=blue, urlcolor=blue]{hyperref}
\usepackage{multirow}
\usepackage{braket}
\usepackage[normalem]{ulem}
\usepackage{multirow}
\usepackage{makecell}
\usepackage{tikz}
\usetikzlibrary{calc}
\usetikzlibrary{shapes.multipart}

\newcommand{\comments}[1]{}

\newcommand{\av}[1]{\langle#1\rangle}

\newcommand{\stkout}[1]{\ifmmode\text{\sout{\ensuremath{#1}}}\else\sout{#1}\fi}

\makeatletter
\def\l@subsubsection#1#2{}
\makeatother

\begin{document}

\title{Resummation-based Quantum Monte Carlo for Entanglement Entropy Computation}

\author{Menghan Song}
\affiliation{Department of Physics and HKU-UCAS Joint Institute of Theoretical and Computational Physics, The University of Hong Kong, Pokfulam Road, Hong Kong SAR, China}

\author{Ting-Tung Wang}
\affiliation{Department of Physics and HKU-UCAS Joint Institute of Theoretical and Computational Physics, The University of Hong Kong, Pokfulam Road, Hong Kong SAR, China}

\author{Zi Yang Meng}
\email{zymeng@hku.hk}
\affiliation{Department of Physics and HKU-UCAS Joint Institute of Theoretical and Computational Physics, The University of Hong Kong, Pokfulam Road, Hong Kong SAR, China}

\begin{abstract}
Based on the recently developed resummation-based quantum Monte Carlo method for the SU($N$) spin and loop-gas models, we developed a new algorithm, dubbed ResumEE, to compute the entanglement entropy (EE) with greatly enhanced efficiency. Our ResumEE exponentially speeds up the computation of the exponentially small value of the $\langle e^{-S^{(2)}}\rangle$, where $S^{(2)}$ is the 2nd order R\'enyi EE, such that the $S^{(2)}$ for a generic 2D quantum SU($N$) spin models can be readily computed with high accuracy. We benchmark our algorithm with the previously proposed estimators of $S^{(2)}$ on 1D and 2D SU($2$) Heisenberg spin systems to reveal its superior performance and then use it to detect the entanglement scaling data of the N\'eel-to-VBS transition on 2D SU($N$) Heisenberg model with continuously varying $N$. Our ResumEE algorithm is efficient for precisely evaluating the entanglement entropy of SU($N$) spin models with continuous $N$ and reliable access to the conformal field theory data for the highly entangled quantum matter.    
\end{abstract}

\date{\today}
\maketitle

\noindent{\textcolor{blue}{\it Introduction.}---}
The SU($N$) spin~\cite{readSome1989,largeRead1991,kawashimaGround2007} and loop-gas~\cite{aletUnconventional2006,charrierPhase2010,Nahum3Dloop2011,NahumPhase2013,NahumDQC2015,sreejithEmergent2019} models, especially at the large-$N$ limit, are theoretically important and have been intensively investigated over the years~\cite{ribhu2008quantum,ChesterEvidence2023,Metlitski2008Monopoles}. From the numerical side, there have been significant developments in the quantum Monte Carlo (QMC) simulation techniques to solve such models~\cite{kawashimaLoop1994,haradaNeel2003,assaadPhase2005,kawashimaGround2007,BeachcontinuousN2009,sandvikLoop2010,quantumKaul2011,latticeKaul2012,possibilityHarada2013}. Based on the valence-bond basis~\cite{sandvikLoop2010}, a few projection QMC works tried to determine the phase transition from the antiferromagnetic (AFM) N\'eel state to the valence bond solid (VBS) state as the $N$ in SU($N$) increases~\cite{BeachcontinuousN2009}. The finite temperature path-integral QMC algorithm has also been developed~\cite{kawashimaLoop1994,haradaNeel2003} and many interesting phase transitions, especially when further interactions beyond the nearest-neighbor have been included, are subsequently discovered~\cite{quantumKaul2011,latticeKaul2012,blockFate2013,possibilityHarada2013}, similar discoveries are also made in the loop-gas models~\cite{aletUnconventional2006,charrierPhase2010,Nahum3Dloop2011,NahumPhase2013,NahumDQC2015,sreejithEmergent2019}. These efforts have culminated in the recent discovery that the deconfined quantum criticality (DQC) in its SU($N$) spin model realization ($J_1$-$J_2$ SU($N$) model on square lattice with spins in the fundamental representation on sub-lattice A and the conjugate representation on sub-lattice B) will be necessarily recovered when $N\geq 8$ from the entanglement entropy (EE) measurement~\cite{song2024deconfined}.

% It is in the work of EE computation in SU($N$) $J_1$-$J_2$ spin model~\cite{songDeconfined2023} that we found either the valence-bond basis projection QMC~\cite{Hastings2010Measuring} or the SSE finite temperature QMC simulations of the SU($N$) model become cumbersome when $N$ is large, with larger auto-correlation time and noisier data compared with their SU(2) or SU(3) cousins. 
The Entanglement entropy, as a non-local quantum measurement, follows the scaling form of "area law"~\cite{calabreseEntanglement2004,fradkinEntanglement2006,laflorencieQuantum2016}. The corrections to the area law usually contain universal information~\cite{metlitskiEntanglement2011,luitzUniversal2015,NicoSpinwave2015,LuitzQMCSU(2)2017,zhaoScaling2022,liaoTeaching2023,song2023quantum,liuDisorder2023} and can be used to identify quantum phases~\cite{Isakov2011,Jiang2012Identifying,MatthewKagome2020,zhaoMeasuring2022} and phase transitions~\cite{zhaoScaling2022,song2023quantum,song2024deconfined,SwingleStructure2012,Inglis_2013,KallinEntanglement2013,liaoTeaching2023,liuDisorder2023} in lattice model simulations. QMC simulations of R\'enyi entanglement entropy offer the ability to probe these fundamental properties in large system sizes and higher dimensions by interpreting the R\'enyi entropies from partition functions of replicated space-time manifolds~\cite{Hastings2010Measuring,Humeniuk2012,luitzImproving2014,luitzUniversal2015,kulchytskyyDetecting2015}. It is in the work of EE computation in SU($N$) $J_1$-$J_2$ spin model~\cite{song2024deconfined}, that we find although the existing EE computation algorithms can efficiently access R\'enyi entropy of SU($N$) spin models at small $N$~\cite{albeOut2017,demidioEntanglement2020,zhaoScaling2022,zhaoMeasuring2022,Bulgarelli2023}, the simulations become cumbersome when $N$ is large, with larger auto-correlation time and noisier data compared with their SU(2) or SU(3) cousins.
% At $T=0$, Hastings
% ${et\ al.}$~\cite{Hastings2010Measuring} managed to compute EE by sampling a “swap”-operator in a projector Monte Carlo
% approach~\cite{SandvikGroundstate2005} of complexity $O(m^{2})$ where $m$ is the projection power. This complexity can be reduced to $O(m)$ using the loop update scheme originally developed for SSE, while this constrains us at SU($N$) spins with integer $N$. Finite temperature algorithms   
The basic problem is that the configuration space in the Monte Carlo sampling process becomes exponentially large as $N$ increases, i.e., each site has $N$ local degrees of freedom, such that traditional stochastic series expansion (SSE) operator loop update~\cite{sandvikLoop2010,Evertzloop2010} has to deal with too many types ($N$ colors) of loops, making it exponentially hard to visit all the configurations, compared with, for example, the SU(2) case. Due to this difficulty, one cannot even compute the local observables, such as the magnetization and VBS order parameters accurately at large-$N$, let alone non-local observables, such as EE where one needs to replicate the partition function, and whose evaluation $\langle e^{-S^{(2)}}\rangle$ where $S^{(2)}$ is the 2nd order R\'enyi EE could render exponential explosion of its variance~\cite{liaoControllable2023,panStable2023}.

\begin{figure}[htp!]
\includegraphics[width=\columnwidth]{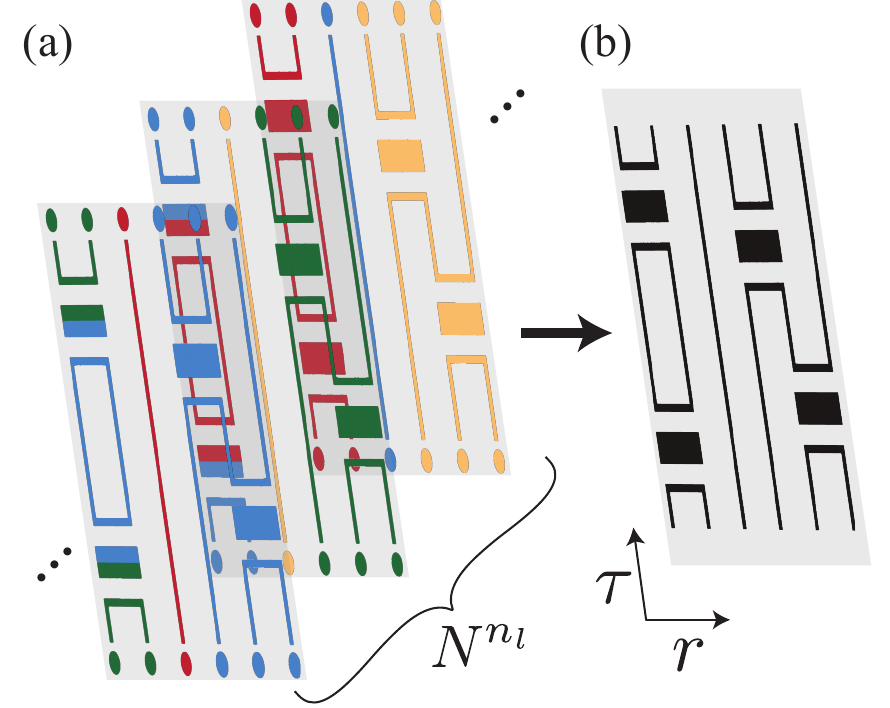}
\caption{\textbf{Concept sketch for the resummation QMC.} In this example, $N^{n_l=4}$ colored spin configurations with the same loop structure (panel (a)) for a general SU($N$) model are resumed into a compact uncolored loop configuration (panel (b)). In the coordinate, $r$ stands for the spatial direction of the lattice, and the $\tau$ is the imaginary time direction in the partition function.}
\label{fig:fig1}
\end{figure}

Apparently, a new QMC updated scheme is needed to reduce the complexity introduced by the multi-color loops in the SU($N$) or loop-gap models, and recently, there appears to be such a scheme that serves the purpose~\cite{desairesum2021}. 
In the resummation-based QMC method, as schematically shown in Fig.~\ref{fig:fig1}, Desai and Pujari introduced the finite-$T$ operator-loop updates which directly handle uncolored loops without any reference to the underlying spin states and, therefore, manage to re-sum all the contributions from numerous colored loops into one single uncolored loop. This method works well in medium-to-large $N$ cases and, more importantly, inspires us to develop a more efficient sampling algorithm for the EE computation in the SU($N$) quantum spin models, dubbed ResumEE -- resummation-based quantum Monte for entanglement entropy computation.

In this work, we develop the ResumEE and demonstrate its capability to tackle generic SU($N$) spin models at $continuous$ $N$. 
%Our algorithm converts the traditional evaluation of the exponentially small value of the $\langle e^{-S^{(2)}}\rangle$, where $S^{(2)}$ is the 2nd order R\'enyi entanglement entropy, to an important sampling process with polynomial accuracy such that the EE $S^{(2)}$ for a generic 2D quantum SU($N$) spin models can be readily computed.
Benefiting from the compact configuration space, we find that not only at large-$N$ but also in the well-studied SU(2) case, our algorithm greatly improves the previous EE computation methods, such as those in Refs.~\cite{luitzImproving2014,luitzUniversal2015,kulchytskyyDetecting2015} in both 1D and 2D cases. Therefore, the ResumEE algorithm successfully mitigates the long-standing issue of the exponential explosion of EE's variance, which has plagued most of the previous EE estimators in QMC~\cite{metlitskiEntanglement2011,assaadEntanglement2014,broeckerNumerical2016,changEntanglement2015,Humeniuk2012,luitzImproving2014,luitzUniversal2015,kulchytskyyDetecting2015}. 

We first benchmark our algorithm with previously proposed estimators of EE~\cite{luitzImproving2014,luitzUniversal2015,kulchytskyyDetecting2015} on 1D and 2D SU($2$) Heisenberg spin systems to reveal its superior performance and then use it to detect the entanglement scaling data of the N\'eel-to-VBS transition on 2D SU($N$) Heisenberg model on a square lattice with $continuously$ varying $N$. Our results reveal that although previous work suggested there is a continuous transition at $N_c \approx 4.57(5)$~\cite{BeachcontinuousN2009}, the more sensitive EE computation suggests that the transition is actually weakly- first-order with a substantial log-correction to the area law scaling in EE when the entanglement region has a smooth boundary. Such a log-correction at a quantum critical point (QCP) violates the conformal field theory (CFT) constraint, and we conjecture, is inherited from the domains that exhibit SU($N$) continuous symmetry-breaking with gapless Goldstone modes, which coexist with the VBS domains at their weakly- first-order transition point. Once pushed to substantially larger system sizes, our local order parameter measurements convey the same message of phase coexistence of the N\'eel and VBS states. 

The ResumEE algorithm, therefore, not only demonstrates its superior performance compared with the previous EE computation schemes but offers a sensitive non-local probe that could distinguish the fundamental information of the transition that is otherwise very difficult for conventional probes. We, therefore, believe our ResumEE makes an essential step towards solving the critical problem of precisely evaluating the quantum entanglement in many-body systems and is of unusual intrinsic interest to a broad audience from condensed matter to quantum information and computation. \\

\noindent{\textcolor{blue}{\it Resummation QMC and traditional EE estimators.}---}%
In the path-integral QMC, traditionally, one can series expand the partition function as
\begin{eqnarray}
    Z(\beta)&=&\text{Tr}\{e^{-\beta H}\}\nonumber\\
    &=&\sum_{n=0}^{\infty}\frac{(-\beta)^n}{n!}\sum_{\{b_m,\mu_m\}}\sum_{\sigma}\langle \sigma |H_{b_1,\mu_1}H_{b_2,\mu_2}\cdots H_{b_n,\mu_n}|\sigma\rangle\nonumber\\
\label{eq:eq1}
\end{eqnarray}
where $\{b_m,\mu_m\}$ identifies the $m$-th operator at bond $b_m$ in the operator string with operator type $\mu_m$ (diagonal or off-diagonal) and $|\sigma\rangle$ is the basis state the operators $H_{b_m, \mu_m}$ are operated upon (usually it is the product state of the on-site $|S^z_i\rangle$). For the SU($N$) spin model, typical configurations of Eq.~\eqref{eq:eq1}, are shown in Fig.~\ref{fig:fig1} (a), one can image the operator loops as closely packed colored loops~\cite{marshallKaul2015}. And since there are $N$ colors, the total number of configurations is exponentially large, i.e., $N^{n_l}$, where $n_l$  is the number of loops. 

The resummation QMC~\cite{desairesum2021} sums over the spin or color indices of these closely packed loops without changing loop connections along the operator string, such that colored configurations that share the same loop connection, as shown in Fig.~\ref{fig:fig1} (a), are compressed into one uncolored loop configuration in Fig.~\ref{fig:fig1} (b). The simplified partition function is now
\begin{eqnarray}
Z(\beta)=\sum_{n=0}^{\infty}\frac{(-\beta)^n}{n!}N^{n_l}\sum_{\{b_n\}}h_{b_1}h_{b_2}\cdots h_{b_n},
\label{eq:eq2}
\end{eqnarray}
where one only needs to record the operator position $b_m$.  In our SU($N$) Hamiltonian, $h_{b_m}=-\frac{J}{N}$ is the spin-symmetric matrix element, and $n_l$ is the number of loops in a given configuration. We emphasize that the uncolored nature of the loops emerges in the presence of SU($N$) symmetry, and the diagonal and off-diagonal operators in Eq.~\eqref{eq:eq1} contribute the same element in Eq.~\eqref{eq:eq2}. Another interesting observation is that the symmetry index $N$ is now a parameter that can be tuned continuously instead of only the integer in its original form.

\begin{figure}[htp!]
\includegraphics[width=\columnwidth]{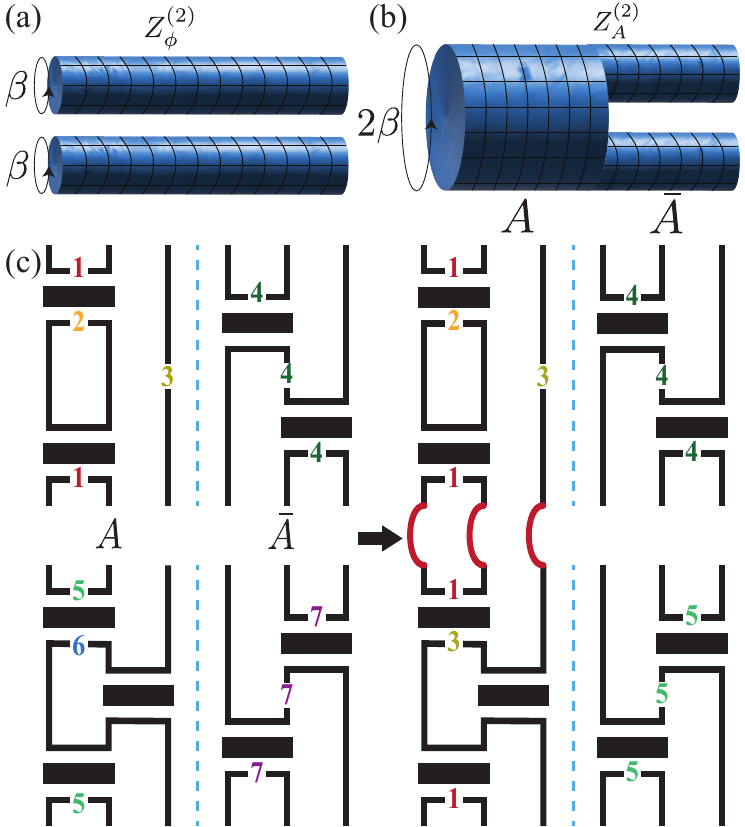}
\caption{\textbf{The resummation-based EE computation.} Panel (a) shows the independent ensemble of two replicas corresponding to the partition function $Z^{(2)}_{\emptyset}$, and panel (b) shows the ensemble where the entanglement region $A$ is glued with imaginary time-periodic boundary condition $2\beta$, corresponding to partition function $Z^{(2)}_{A}$. The environment $\overline{A}$ still has the boundary condition of $\beta$. Panel (c) shows an example of measuring $\av{P_{\emptyset \rightarrow A}}$ in ResumEE algorithm. \textcolor{black}{Loops labeled with the same index are those connected. The left and right parts of the panel (c) have the imaginary time boundary conditions illustrated in panel (a) and panel (b), respectively.}}
\label{fig:fig2}
\end{figure}

Based on such a resummation scheme of the partition function of the SU($N$) Hamiltonian in Eq.~\eqref{eq:eq2}, we can discuss the computation of the entanglement entropy based on the evaluation of the partition function ratio on different space-time manifolds~\cite{calabreseEntanglement2004, Humeniuk2012,luitzImproving2014}. Here, we use the second order R\'enyi EE as an example. As shown in Fig.~\ref{fig:fig2}, for a quantum many-body system, the EE with entanglement region $A$ can be expressed as
\begin{equation}
S^{(2)}_A=-\ln{\frac{Z^{(2)}_A}{Z^{(2)}_\emptyset}}=-\ln{\frac{\av{P_{\emptyset\rightarrow A}}}{\av{P_{A\rightarrow \emptyset}}}},
\label{eq:eq3}
\end{equation}
and the two partition functions $Z^{(2)}_\emptyset$ and $Z^{(2)}_A$ are computed on the two different space-time manifolds in Fig.~\ref{fig:fig2} (a) and (b), respectively. The last equity in Eq.~\eqref{eq:eq3} is achieved because %in an equilibrium Monte Carlo process, one shall have the detailed balance $W_{\emptyset}P_{\emptyset\rightarrow A}=W_{A}P_{A\rightarrow\emptyset}$ where $\langle P_{\emptyset\rightarrow A}\rangle$ ($\av{P_{A \rightarrow \emptyset}}$) is the probability of configuration satisfying the boundary condition of $Z_A$ ($Z_\emptyset$) when carrying out the simulation in $Z_\emptyset$ ($Z_A$)~\cite{luitzImproving2014,luitzUniversal2015}. 
we define the probabilities $\langle P_{\emptyset\rightarrow A}\rangle$ and $\langle P_{A\rightarrow \emptyset}\rangle$ as  
\begin{eqnarray}
\langle P_{X\rightarrow \bar{X}}\rangle &=&\frac{Z^{(2)}_{X\cap \bar{X}}}{Z^{(2)}_{X}}=\frac{\sum_{C\in X\cap \bar{X}}W(C)}{\sum_{C\in X}W(C)} \nonumber\\
&=&\frac{\sum_{C\in X}W(C)\delta_{\bar{X}}(C)}{\sum_{C\in X}W(C)}
=\langle\delta_{\bar{X}}(C)\rangle_X\text{,}
\label{eq:eq4}
\end{eqnarray}
and 
\begin{equation}
\delta_X(C)= 
\begin{cases}
    1,& \text{if } C\in X\\
    0,& \text{if } C\notin X
\end{cases}\text{,}
\end{equation}
where $X$ and $\bar{X}$ are dummy variables being $\emptyset$ and $A$ interchangeably and $W(C)$ is the configurational weight for configuration $C$ specified in the partition function Eq.~\eqref{eq:eq2}. 

In Eq.~\eqref{eq:eq4}, when $C\in\emptyset$, the Monte Carlo average is carried out in the space-time manifolds such that the loop configurations are on the two independent replicas as in Fig.~\ref{fig:fig2} (a); and when $C\in A$, the loop configurations are on the manifold with the entanglement region $A$ connected/glued but the environment $\overline{A}$ are independent, as in Fig.~\ref{fig:fig2} (b). For computing $\langle P_{\emptyset \rightarrow A} \rangle$, the $C\in \emptyset\cap A$ in the numerator in Eq.~\eqref{eq:eq4} means that although we sample in the two independent replicas, i.e., $C\in\emptyset$, we need to count the spin configurations that satisfy the periodic boundary condition of the glued manifold of $C\in A$, hence the $\delta_A$ function; and vice versa for $\langle P_{A \rightarrow \emptyset}\rangle$.

In the traditional path-integral QMC estimator of EE~\cite{Humeniuk2012,luitzImproving2014}, one measures $\delta_X$ by comparing the spins in $A$ at the imaginary time boundaries. If all the spins satisfy the boundary condition of $X$, a 1 is recorded, and 0 otherwise. As already indicated in Ref.~\cite{Humeniuk2012}, such a counting method becomes inefficient when the entanglement region $A$ is large since the ratio becomes exponentially small and the MC process becomes a sampling of rare events. Although later an improved version is introduced in Ref.~\cite{luitzImproving2014} taking advantage of symmetry, measuring an exponentially small quantity by sampling 1s and 0s is still ill-conditioned, as one would need roughly $e^{S_A^{(2)}}$ MC steps to get an average value of order $e^{-S_A^{(2)}}$. The computation cost scales exponentially with $S_A^{(2)}$ and hence the linear system size in 2D, as the leading term in EE is the area law $S_A^{(2)}\sim L$. \\

\noindent{\textcolor{blue}{\it The ResumEE algorithm.}---}%
In the ResumEE, the situation is different. Here, one can effectively sample $N^{n_l}$ spin configurations with each loop configuration. We can, therefore, sample a small but finite value of $\langle P_{A \rightarrow \emptyset}\rangle$ and $\langle P_{\emptyset \rightarrow A} \rangle$ each time instead of waiting for a 1 to appear indefinitely. To be more specific, the two probabilities $\langle P_{A \rightarrow \emptyset}\rangle$ and $\langle P_{\emptyset \rightarrow A} \rangle$ can be evaluated as the ratios of numbers of possible spin configurations under different boundary conditions. They can be measured with the reduction of the number of loops when one enforces the additional boundary condition, as the number of loops determines the number of possible spin configurations. That is,
\begin{equation}
P_{\emptyset \rightarrow A}=\frac{N^{n_{glued}}}{N^{n_{1}}N^{n_{2}}}
=N^{n_{glued}-(n_1+n_2)},
\label{eq:eq6}
\end{equation}
where $n_1$ and $n_2$ are the numbers of loops in each independent replica, and $n_{glued}$ is the number of loops when one glued them together in region $A$.

Fig.~\ref{fig:fig2} (c) explicitly shows an example of measuring $P_{\emptyset \rightarrow A}$ according to Eq.~\eqref{eq:eq6}. Before gluing the loops into the $C\in A$ manifold, one has $n_1+n_2=7$. After applying the additional boundary condition, one has $n_{glued}=5$. Therefore, one sample a $P_{\emptyset \rightarrow A}=N^{-2}$ with this loop configuration, as opposed to sample a $1$ every $N^{2}$ times if one were to measure with the original method with spin configurations. $P_{A \rightarrow \emptyset}$ can be measured in a similar manner, with the only difference being that the additional boundary condition is cast on the top and bottom of one of the replicas.

In practice, one only needs to count the number of distinct groups reduced at the boundary of imaginary time during such a gluing process, as the loops in the middle of time propagation will appear in both cases and cancel out in the ratio of partition functions, e.g., loops 2, 4 and 7 in Fig.~\ref{fig:fig2} (c). Since the number of spin configurations measured in the resummation algorithm increases exponentially with system size, the ResumEE succeeded in mitigating the exponential decay in EE, or the exponential explosion of its variance~\cite{liaoControllable2023,panStable2023,zhangIntegral2024}, as it happened in the previous methods~\cite{luitzImproving2014,luitzUniversal2015,kulchytskyyDetecting2015}. We now turn to the results to demonstrate such superior performance.\\

\noindent{\textcolor{blue}{\it Benchmark with SU(2) Heisenberg model.}---} \textcolor{black}{We consider the quantum spin model with Hamiltonian 
\begin{equation}
H=-J\sum_{\langle i,j \rangle}P_{ij},
\label{eq:H_hsb}
\end{equation}
where $\langle i,j \rangle $ denotes the nearest neighbor interaction. $P_{ij}$ is the SU($N$) singlet projector, defined as $P_{ij}=\frac{1}{N}\sum_{\alpha,\beta}|\alpha_i\alpha_j\rangle\langle \beta_i\beta_j|$ with $\alpha,\beta$ ranging from 1 to $N$. }The lattice is $d=(1,2)$ dimension bipartite (in these cases, chain and square lattice) lattice such that the $\alpha$ spins are in the fundamental representation on one sub-lattice and $\beta$ spins are in the conjugate representation on the other sub-lattice. 

\begin{figure}[htp!]
\includegraphics[width=\columnwidth]{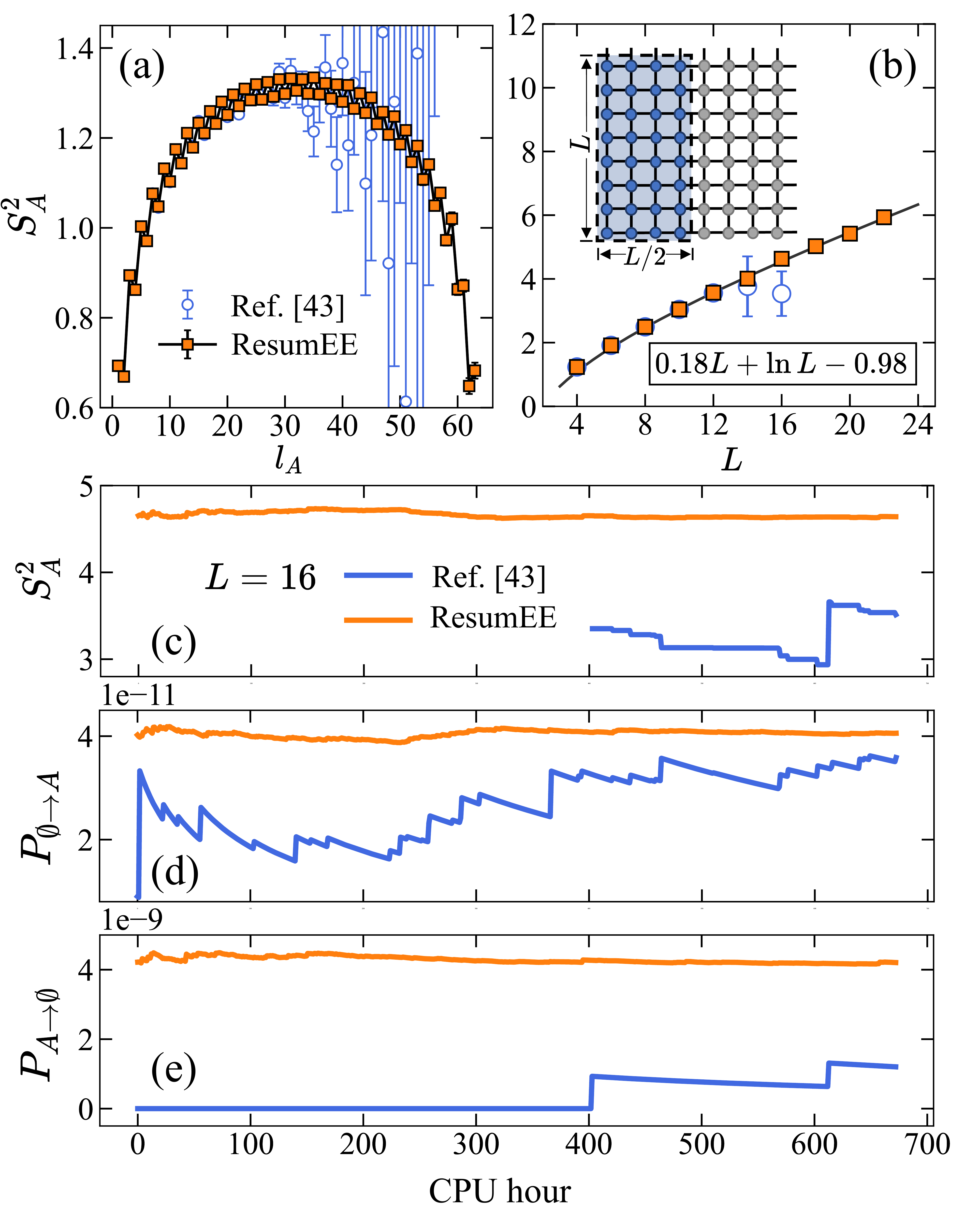}
\caption{\textbf{Benchmark $S^{(2)}_A$ for 1D and 2D SU(2) Heisenberg model.} (a) Comparison of our ResumEE method and previous EE estimator in Ref.~\cite{luitzImproving2014} on a 1D Heisenberg chain with $L=64$ and inverse temperature $\beta=128$. The size of the entanglement region is $l_A$. (b) Comparison of ResumEE and EE estimator in Ref.~\cite{luitzImproving2014} on 2D SU(2) square lattice with $L=4,6,...,16$ and $\beta=L$. The entanglement region $A$ is of size $L\times L/2$ with a smooth boundary of length $l_A=2L$. In all cases, we perform the same amount of CPU hours for fair comparison. In (a), it is obvious that when $l_A > L/2$, the traditional EE results experience the exponential explosion of variance, whereas the ResumEE has controlled errors. In (b), only the ResumEE can yield the expected scaling of $S_A^{(2)}(L)=a L + b\ln L +c$, where $b=\frac{N_G}{2}=1$ with $N_G=2$ the number of Goldstone modes in the N\'eel phase of AFM SU(2) Heisenberg model on square lattice. Panels (c) (d) and (d) demonstrate the convergence of $S^2_A$, $P_{\emptyset\to A}$ and $P_{A\to \emptyset}$ against simulation time at $L=16$ in panel (b), respectively.}
\label{fig:fig3}
\end{figure}

To benchmark the performance of the ResumEE algorithm, we first compute the second order R\'enyi EE of two well-known examples at SU(2) cases, the spin-1/2 Heisenberg chain, and the spin-1/2 square lattice AFM Heisenberg model. The results are shown in Fig.~\ref{fig:fig3}. One sees in the 1D case (Fig.~\ref{fig:fig3} (a)) that the ResumEE always has controlled errorbars as $l_A$ (the entanglement region) increases, while the traditional EE estimators in Ref.~\cite{luitzImproving2014} quickly experience the exponential explosion of the variance~\cite{luitzImproving2014,luitzUniversal2015,kulchytskyyDetecting2015}. The contrast is even clearer in the 2D case in Fig.~\ref{fig:fig3} (b). In this case, the square lattice SU(2) antiferromagnetic model has spontaneous symmetry breaking N\'eel phase with $N_G=2$ Goldstone modes at ground state, and since the entanglement region acquires smooth boundary (see the inset of Fig.~\ref{fig:fig3} (b)), one expects the scaling behavior of $S_A^{(2)}(L)=a L + b\ln L +c$ with the leading area law and the universal coefficient $b=\frac{N_G(d-1)}{2}=1$ in the sub-leading term~\cite{metlitskiEntanglement2011,kulchytskyyDetecting2015,demidioEntanglement2020,zhaoMeasuring2022,panStable2023} with $d=2$ the spatial dimension. One sees that the ResumEE readily captures the expected scaling behavior, but the traditional EE estimators cannot offer precise values for $L=14,16$, and cannot sample a value for $L>16$ for the CPU hour we used since none of $\langle P_{A \rightarrow \emptyset}\rangle$ or $\langle P_{\emptyset \rightarrow A} \rangle$ measurements find spin configurations with satisfactory boundary conditions. At $L=16$, one can see in panel (c)-(e) that the measurables obtained in ResumEE converge much faster than those sampled using the traditional method in Ref.~\cite{luitzImproving2014}. The traditional method actually failed to sample any effective value of $P_{A\to\emptyset}$ until 400 CPU hours (Fig.~\ref{fig:fig2} (e)), so no EE data was recorded (Fig.~\ref{fig:fig2} (c)). To make a fair comparison, we have purposely kept the same amount of CPU hours for the ResumEE and traditional EE estimator in all the data of Fig.~\ref{fig:fig3}.\\

\begin{figure}[htp!]
\includegraphics[width=\columnwidth]{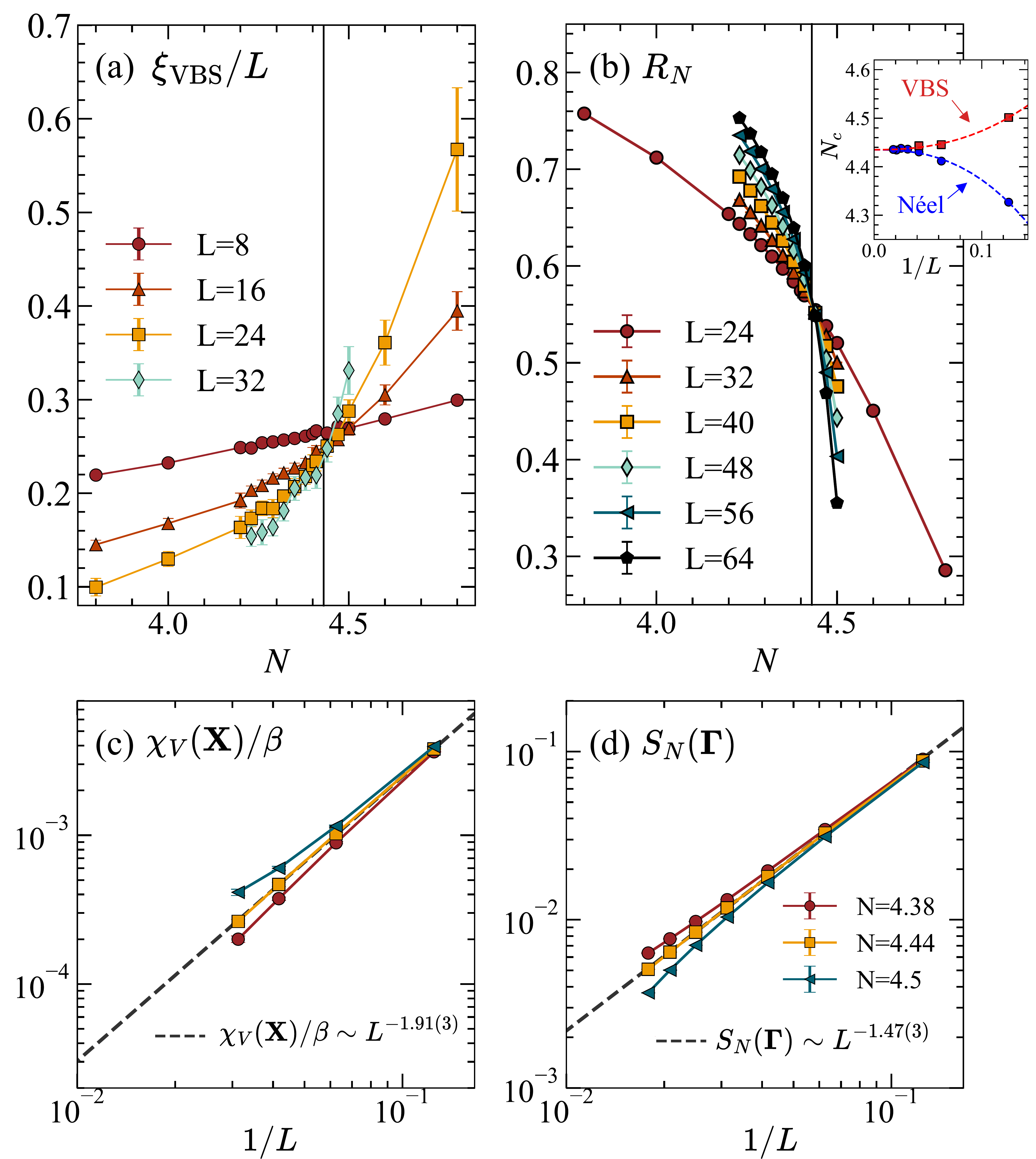}
\caption{\textcolor{black}{\textbf{Determine N\'eel-VBS transition in the SU($N$) model from local probes.} Panels (a) and (b) demonstrate the cross of VBS correlation length and N\'eel order correlation ratio, respectively. The inset in (b) shows the convergence to a common critical point of the two crossings as $L\to\infty$ and determines $N_c$. Panels (c) and (d) demonstrate how the finite-size VBS and N\'eel order parameters vary as the system size inside two phases and near the transition point. Up to the present system sizes, one can see the scaling dimensions of the two order parameters are different.}}
\label{fig:fig4}
\end{figure}

\noindent{\textcolor{blue}{\it Phase transition in the SU($N$) Heisenberg model with continuously varying $N$.}---}
The more distinctive feature of ResumEE is its ability to compute EE efficiently at non-integer $N$ values. Then, we turn to the quantum phase transition in the SU($N$) model as a function of continuously varying $N$. It is known that this model has a N\'eel ordered ground state at SU($N\le 4$) while a VBS ground state at SU($N\ge 5$)~\cite{kawashimaGround2007,haradaNeel2003,kawashima2004}. Later, Ref.~\cite{BeachcontinuousN2009} with projection QMC in the VBS basis suggested a continuous phase transition at $N_c\approx4.57(5)$ separating the N\'eel phase with onsite spin rotational symmetry breaking and the VBS phase with $Z_4$ lattice symmetry breaking. This finding is interesting because the transition, if continuous, shares the flavors of deconfined quantum criticality (DQC) in that two spontaneously symmetry-breaking phases meet at a single critical point with one tuning parameter. Although we have to acknowledge that the present case is different from the DQC in that the symmetry of the Hamiltonian changes when $N$ changes,  the N\'eel-VBS transition in the present case could nevertheless offer new ingredients of novel quantum phase transitions.

However, previous analysis of the continuous phase transition~\cite{BeachcontinuousN2009} assumes that both the N\'eel and VBS phases share the same anomalous dimension, i.e., $\eta_{N}=\eta_{V}$. It was reported later that $\eta_N$ deviates from $\eta_V$ in the SU($N$) DQC scenario with $N>2$, and $\eta_{N}=\eta_{V}$ is a special case for SU($2$)~\cite{latticeKaul2012}. Although our case is not identical to the DQC scenario, it is reasonable to investigate the critical $N_c$ more carefully without a priori assumption such as $\eta_{N}=\eta_{V}$. Also, the previous results are based on the projection QMC at the ground state in which the projection process holds an $O(m^2)$ complexity at non-integer $N$ values, where the projection power $m$ is usually $m\sim L^2$~\cite{sandvikLoop2010,SandvikGroundstate2005}.
With the resummation QMC in the path-integral whose complexity near QCP is $O(\beta L^2)$~\cite{desairesum2021}, we can simulate larger system sizes (up to $L=64$ and inverse temperature $\beta=L$) than those in Ref.~\cite{BeachcontinuousN2009}. So, we decided to examine the N\'eel-VBS transition point in a more accurate manner.

\textcolor{black}{We solve the problem from two different directions. First, we adopt the traditional practice by looking into the local order parameters for N\'eel and VBS phases and use the finite-size scaling analysis to locate the more accurate critical region and the transition point. These results are shown in Fig.~\ref{fig:fig4}. We define the SU($N$) N\'eel correlation function $C_N(\mathbf{r},\mathbf{r^{\prime}})$ to be the expectation value of the projection operator, $\frac{1}{L^4}{\textstyle \sum_{\mathbf{r,r^{\prime}}}^{}} P_{\mathbf{r,r'}}$. The spin structure factor $S_N$ is the Fourier transformation of $C_N$. Since the Hamiltonian Eq.~\eqref{eq:H_hsb}  is written in the fundamental representation,  $S_N(\mathbf{q})$ should peak at ${\mathbf{\Gamma}=(0,0)}$ inside the N\'eel phase.  Similarly, we define SU($N$) VBS correlation function as $C_V(\mathbf{r})=\int_{0}^{\beta}d\tau\langle P_{\mathbf{0},\mathbf{0+\hat{e}}_x}(0)P_{\mathbf{r},\mathbf{r+\hat{e}}_x} (\tau) - \langle P_{\mathbf{0},\mathbf{0+\hat{e}}_x}(0) \rangle ^{2}\rangle $ in $x$ lattice direction (and analogously for $y$ direction). $C_V(\mathbf{r})$ can be estimated as $\langle n(b_1)n(b_2)-\langle n(b)\rangle^2-\delta_{b_1,b_2}n(b_1)\rangle/\beta$, with $n(b)$ denoting the number of projection operators acting on bond $b$, and $b_1, b_2$ are two bonds separating with distance $r$ is $x$ direction. Then, the VBS susceptibility $\chi_{\mathrm{VBS}}$ is the Fourier transformation of $C_V$, which should develop a peak at $\mathbf{X}=(\pi,0)$ inside the VBS phase. Typical data of the above quantities in both phases are given in Appendix~\ref{sec:appB}.}

\textcolor{black}{To locate the $N_c$ at which the transition happens, we track the crossing point of the N\'eel correlation ratio $R_N$ and $\xi_{\mathrm{VBS}}/L$   at different system sizes, where $\xi_{VBS}$ denotes the VBS correlation length~\cite{latticeKaul2012,jin2013thermal}, 
\begin{equation}
R_N=1-\frac{S_N(\mathbf{\Gamma}+ \mathbf{q_1})}{S_N(\mathbf{\Gamma)}},\ \xi_{\mathrm{VBS}}=\frac{L}{2\pi}\sqrt{\frac{\chi_V(\mathbf{X})}{\chi_V(\mathbf{X+q_2})}-1}.
\label{eq:corr_ratio}
\end{equation}
Here, we choose $\mathbf{q_1}=(2\pi/L,0)$ and $\mathbf{q_2}=(4\pi/L,0)$ for better data quality and note that we only keep $\chi_{\mathrm{VBS}}$ data up to $L=32$ because measurements of $C_V$ become too noisy at large $N$ and $L$. One sees from Fig.~\ref{fig:fig4} (a) and (b) that both the dimensionless quantity $R_N$ and $\xi_{\mathrm{VBS}}/L$ have crossing points for various system sizes. A power law fitting of crossings for adjacent system sizes eventually points to a single point, $N_c = 4.43(2)$ at the thermodynamic limit, $1/L= 0$, as shown in the inset of Fig.~\ref{fig:fig4} (b). A critical point near $N_c\approx 4.43$ can also be seen from the scaling of order parameters as a function of system sizes. One can see from Fig.~\ref{fig:fig4} (c) and (d) that at $N=4.44\approx N_c$, both $\chi_V(\mathbf{X})/\beta$ and $S_N(\mathbf{\Gamma})$ vanish in power law of system sizes, indicating a phase transition, although we note the exponents in the power law are different, and it means that the in this case, $\eta_N \neq \eta_V$. Although there is no lucid indication for a first-order transition, it is worth suspecting a weakly-first-order transition analogous to several DQC scenarios at small $N$s~\cite{song2024deconfined,zhaoScaling2022,possibilityHarada2013,latticeKaul2012}. Distinguishing a weakly first-order transition from a continuous one is a formidable task. It can not definitely be answered with the measurement of local observables, even though our $L$ is already larger than the previous work.}

\begin{figure}[htp!]
\includegraphics[width=\columnwidth]{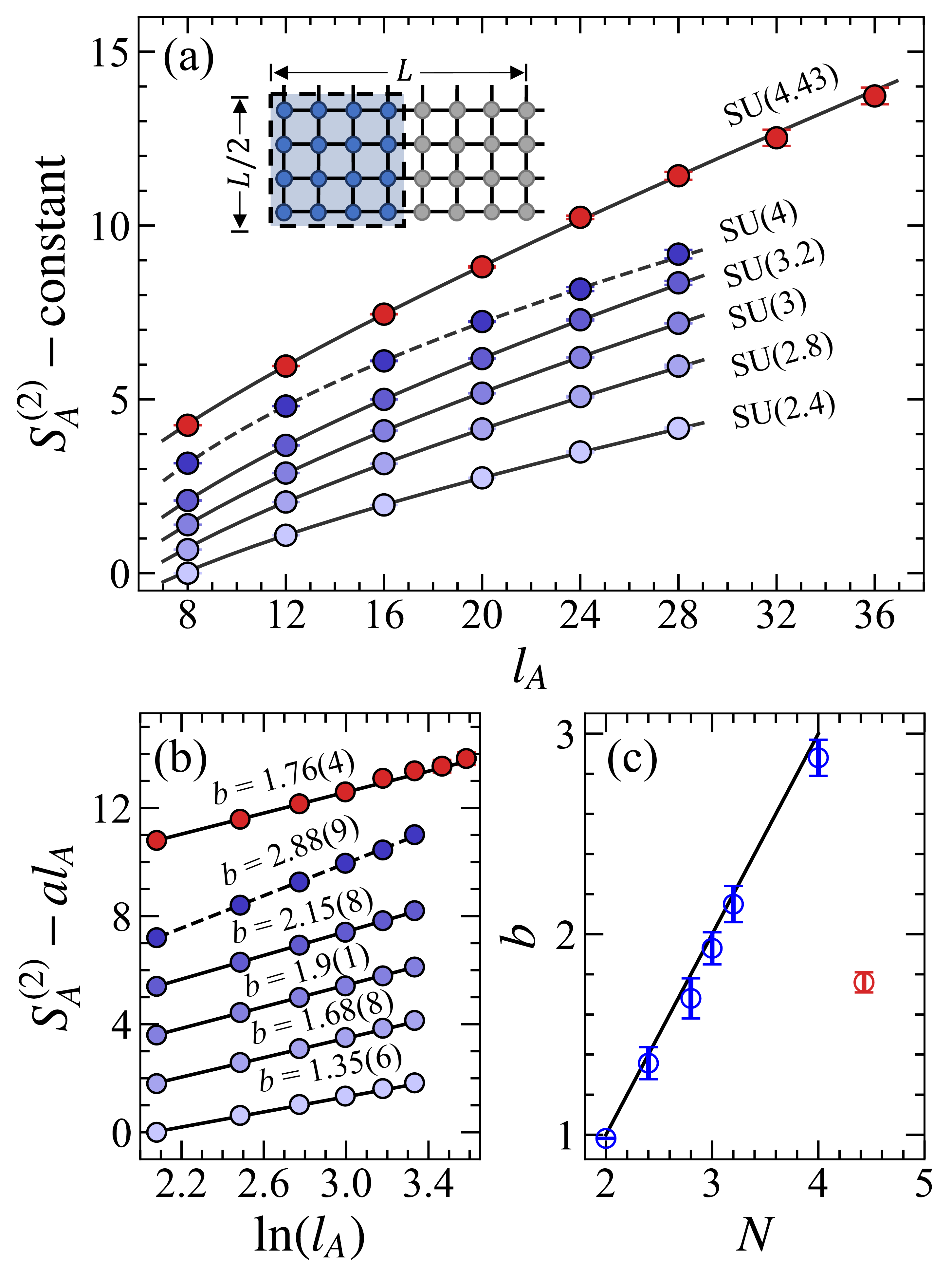}
\caption{\textbf{Scaling of entanglement entropy at different $N$ in the SU($N$) model.} Panel (a) shows $S^2_A(l_A)$ obtained from the ResumEE method with smooth boundary at the transition point $N_c=4.43$ determined from the Binder ratio crossings in Fig.~\ref{fig:fig4}. A clear sub-leading logarithmic correction is seen with the \textcolor{black}{coefficient $b=1.76(4)$} (panel (b)). We also show the EE for SU(2.4), SU(2.8), SU(3), SU(3.2) and SU(4), these systems are well inside the N\'eel phase with $N_G=2(N-1)$ Goldstone modes, such that their sub-leading logarithmic corrections yields $b=\frac{N_G}{2}=N-1$, with $b=1.35(6),1.68(8),1.9(1),2.15(8),2.88(9)$ for the SU(2.4), SU(2.8), SU(3), SU(3.2) and SU(4) cases in panel (b). Note that for SU($4$), we consider the $J_1-J_2$ Hamiltonian (dashed line in (a) and (b)) to enhance the N\'eel order without changing universal information. Panel (c) summarizes the obtained $b$ with respect to $N$, and the black line is $b=N-1$ according to the Goldstone modes inside the SU(N) symmetry-breaking phase. The red dot is the $b$ for SU(4.43), whose finite value suggests the coexistence of the N\'eel state with the VBS state at the transition point. Note that $b$ at $N=2$ adopted from Ref.~\cite{song2024extracting} is also presented in panel (c).}
\label{fig:fig5}
\end{figure}

This is where the non-local EE measurement comes to the rescue. We study the scaling behavior of the $S_A^{(2)}$ at the transition point, $N_c = 4.43$ using ResumEE, and we utilize the incremental trick by gradually increasing the entanglement region to improve the data quality~\cite{Hastings2010Measuring,zhaoScaling2022,zhaoMeasuring2022,panStable2023,liaoControllable2023}. Detailed implementation of the incremental protocol is described in Appendix~\ref{sec:appB}. First, We bi-partition systems with size $L\times L/2$ such that the entanglement region $A$ acquires a smooth boundary with length $l_A=L$ and keep $\beta=L$. For a continuous phase transition described by a unitary CFT, there should be no log-correction to the leading area law at the QCP with a smooth boundary~\cite{metlitskiEntanglement2011,HelmesEEbilayer2014}. In Fig.~\ref{fig:fig5}, we can clearly see for $N=4.43$ a large and \textcolor{black}{positive log coefficient $b=1.76(4)$}, deviating from the CFT constraint. With current knowledge, such log-correction emerged in smooth boundary cut could come from spontaneous continuous symmetry broken with $b=\frac{N_G(d-1)}{2}$. Therefore, such a finite $b$ observed at $N_c=4.43$ may point towards a weakly-first-order phase transition with the co-existence of both N\'eel and VBS order parameters, which is hard to detect at finite system sizes with local observables. 

In order to strengthen our argument, we also investigate the EE scaling inside the N\'eel phase, $N<N_c$ using ResumEE. 
Note that as $N$ approaches $N_c$,  the finite-size effect in EE scaling becomes stronger. \textcolor{black}{We introduce an extra next-nearest neighbor $J_2$ term, $-J_2 \sum_{\langle\langle i j\rangle\rangle} \Pi_{ij} $ to Eq.~\eqref{eq:H_hsb}  at SU(4) case to reduce the finite-size effect,  where $\Pi_{ij}=\frac{1}{N}\sum_{\alpha,\beta} | \beta_i\alpha_j \rangle \langle\alpha_i\beta_j|$  is the SU($N$) generalization of the next-nearest-neighbor ferromagnetic interaction and, therefore, enhances the N\'eel order while keeping universal information, such as the number of Goldstone modes, $N_G$, intact.} In SU(4) case (shown as a dashed line in Fig.~\ref{fig:fig5}), we set $J_2=3.5$~\cite{demidioEntanglement2020}. As shown in Fig.~\ref{fig:fig5} (b), staying deep inside the N\'eel phase, we successfully detect the correct number of Goldstone modes $N_G=2(N-1)$ at SU(3) and SU(4) from EE scaling with the log coefficient, $b=1.9(1)$ and $2.88(9)$ respectively. More interestingly, our ResumEE results find the consistency of $N_G=2(N-1)$ at non-integer $N$ cases, $N=2.4,2.8,3.2$ with $b=1.35(6),1.68(8),2.15(8)$ respectively, as shown in Fig.~\ref{fig:fig5} (c). This is another novel result from ResumEE that the CFT prediction of the log-correction can be analytically continued to non-integer $N$s, or in other words, the number of Goldstone modes $N_G$ can also be fractionalized. Therefore, with the observation inside the N\'eel ordered phase,  the finite log-correction at $N_c$ serves as a clue of continuous symmetry-breaking right at this point, eventually suggesting a weakly-first-order phase transition~\cite{deng2024diag,zhao2024epjq}. 

We also note that the system sizes in Fig.~\ref{fig:fig5} are smaller than those of the order parameter measurements in Fig.~\ref{fig:fig4}. This again shows that the non-local measurement, such as the EE and disorder operator~\cite{wangScaling2022,liuDisorder2023,liuFermion2023,jiangMany2023}, are more sensitive in probing the subtle yet fundamental information of the nature of the phase transitions.
\\

\noindent{{\textcolor{blue}{\it Discussion.}}---}%
The precise computation of the EE for quantum many-body systems is the long-standing obstacle that blockades our full-fledged exploration of the fundamental entanglement and CFT information of quantum matter. In the continuous efforts over the past two decades, various estimators of EE all suffered from the exponential explosion of its variances when the entanglement region just starts to  scale~\cite{metlitskiEntanglement2011,assaadEntanglement2014,broeckerNumerical2016,changEntanglement2015,Humeniuk2012,luitzImproving2014,luitzUniversal2015,kulchytskyyDetecting2015,panStable2023,liaoControllable2023}. In this work, based on the recently developed resummation-based quantum Monte Carlo method for the SU($N$) spin and loop-gas models~\cite{desairesum2021}, we develop a new algorithm -- ResumEE -- manage to mitigate this long-standing obstacle. Our ResumEE exponentially speeds up the computation of the exponentially small value $\langle e^{-S^{(2)}_A}\rangle$ such that the EE for a generic quantum SU($N$) spin models at any continuous $N$ values can be readily computed.

We demonstrate the superior performance of our algorithm over the previous estimators of $S^{(2)}_A$ on 1D and 2D SU($2$) Heisenberg spin systems. More importantly, we use it to detect the entanglement scaling data of the N\'eel-VBS transition on 2D SU($N$) Heisenberg model on a square lattice with $continuously$ varying $N$. Our EE results reveal that although previous work suggested there is a continuous transition at $N_c \approx 4.57(5)$~\cite{BeachcontinuousN2009}, the transition deviates from a CFT description with a substantial log-correction to the area law scaling in EE when the entanglement region has a smooth boundary. Such log-correction may come from the domains that exhibit SU($N$) continuous symmetry-breaking with gapless Goldstone modes, which coexist with the VBS domains at their weakly-first-order transition. Our local order parameter measurements convey the same message of phase coexistence of the N\'eel and VBS states once pushed to much larger system sizes. Our ResumEE also reveals a non-trivial phenomenon: the log-corrections of EE inside the N\'eel phase at non-integer $N$ also follow the CFT prediction, leading to the fractionalization of Goldstone modes. The ResumEE algorithm, therefore, makes an essential step towards extracting the quantum entanglement data and understanding the novel quantum criticality and encourages more works in many-body systems with SU($N$) symmetry at non-integer $N$ values.  \\

\noindent{\it{Acknowledgments.---}}%
We acknowledge the inspiring discussions with Jiarui Zhao, Gaopei Pan, Chengkang Zhou, Bin-Bin Chen and Yuan Da Liao. We thank the support from the
Research Grants Council (RGC) of Hong Kong Special Administrative Region (SAR) of China (Projects
Nos. 17301420, 17301721, AoE/P-701/20, 17309822
and HKU C7037-22G), the ANR/RGC Joint Research
Scheme sponsored by the RGC of Hong Kong SAR of
China and French National Research Agency (Project No. A\_HKU703/22) and the HKU Seed Funding for Strategic Interdisciplinary Research “Many-body paradigm in quantum moir\'e material research”. The authors also acknowledge the Beijng PARATERA Tech CO., Ltd. (URL https://
cloud.paratera.com), the Tianhe-II platform at the National Supercomputer Center in Guangzhou, the HPC2021 system under the Information Technology Services and the Blackbody HPC system at the Department of Physics, University of Hong Kong, for their technical support and generous allocation of CPU time.

\section{Appendix}
\textcolor{black}{In the appendix, we demonstrate qualitatively the speedup of the ResumEE algorithm in Sec.~\ref{sec:appA}. QMC data for order parameters involved in the N\'eel-VBS transition discussed in the main text are exemplified in Sec.~\ref{sec:appB}. Sec.~\ref{sec:appc} introduces the incremental trick used when computing the EE. To strengthen our conclusion, we systematically check the robustness of the anomalous log-correction detected at $N_c$ in Sec.~\ref{sec:appD}.}

% \newpage
% \newpage
% \clearpage
%\onecolumngrid

\appendix
\makeatletter
\setcounter{secnumdepth}{3}	

% {\centering{\bf \uppercase{Supplementary Materials for "Resummation-based Quantum Monte Carlo for Entanglement Entropy Computation"}}}

\section{Performance of ResumEE}
\label{sec:appA}

\begin{figure}[htp!]
\includegraphics[width=\columnwidth]{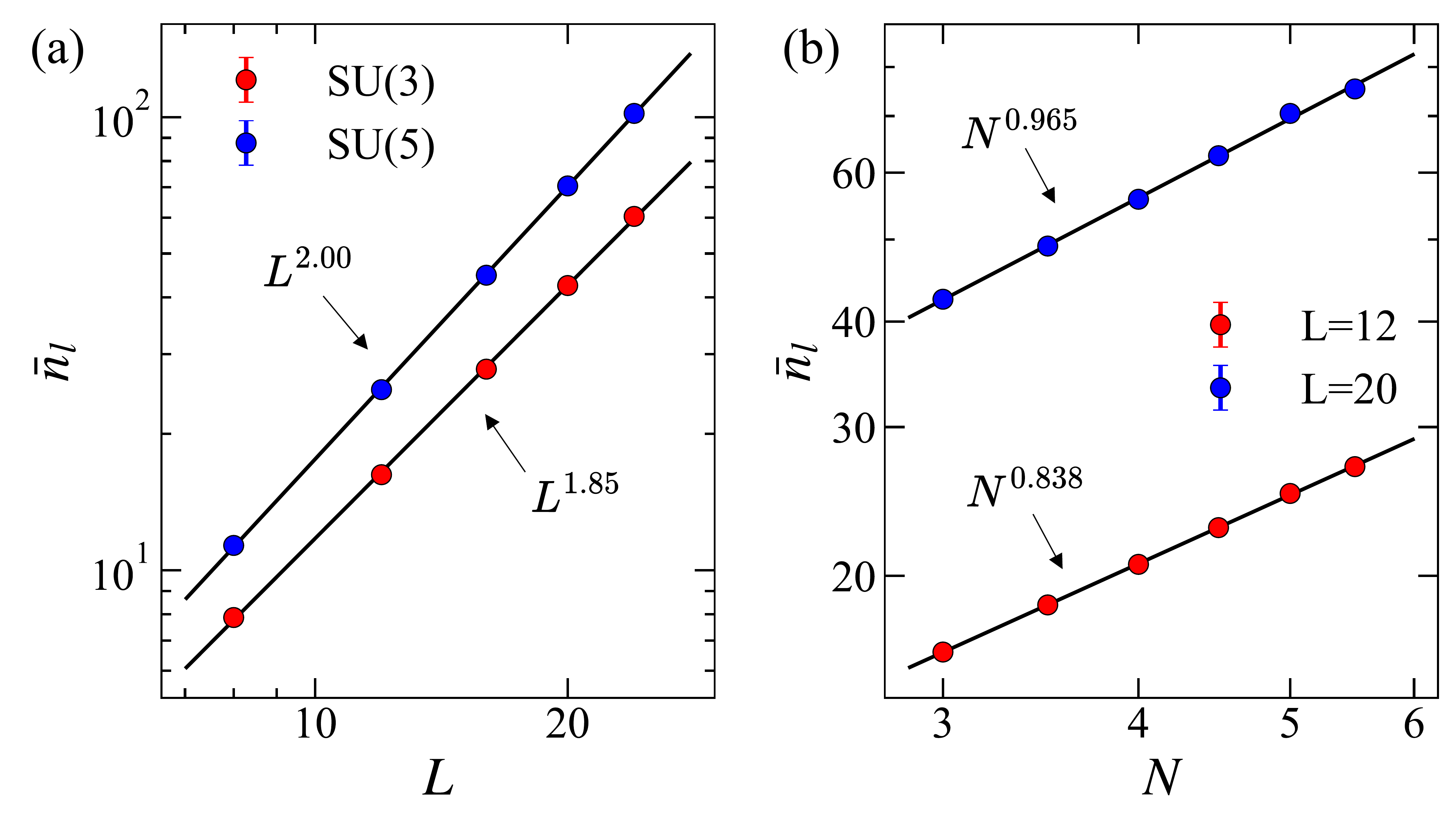}
\caption{\textbf{$\bar n_l$ as a function of (a) $L$ and (b) $N$}. $\bar n_l$ establishes a perfect power-law scaling against the linear system $L$ and $N$, demonstrating the exponential speedup of ResumEE.}
\label{fig:nl_vs_betaN}
\end{figure}

\textcolor{black}{As mentioned in the main text, ResumEE method exponentially improves the sampling efficiency of the traditional entanglement entropy estimator $P_{\emptyset \to A}$ (and $P_{A \to \emptyset}$) through its ability to compress $N^{n_l}$ configurations into one compact, uncolored loop configuration, where $n_l$ denotes the number of loops in an SSE configuration. To be more specific, only those loops at the imaginary time edges contribute to the evaluation of $P$. Let $\bar n_l$ denote the number of loops at the imaginary time edge. It is shown explicitly in Fig.~\ref{fig:nl_vs_betaN}  that $\bar n_l(L, N)$ holds a power law relation with the linear system size $L$ (we keep $\beta=L$ to approach the ground state) and $N$, which demonstrates the exponential speedup of ResumEE.}

\section{QMC data for order parameters}
\label{sec:appB}

\begin{figure}[htp!]
\includegraphics[width=\columnwidth]{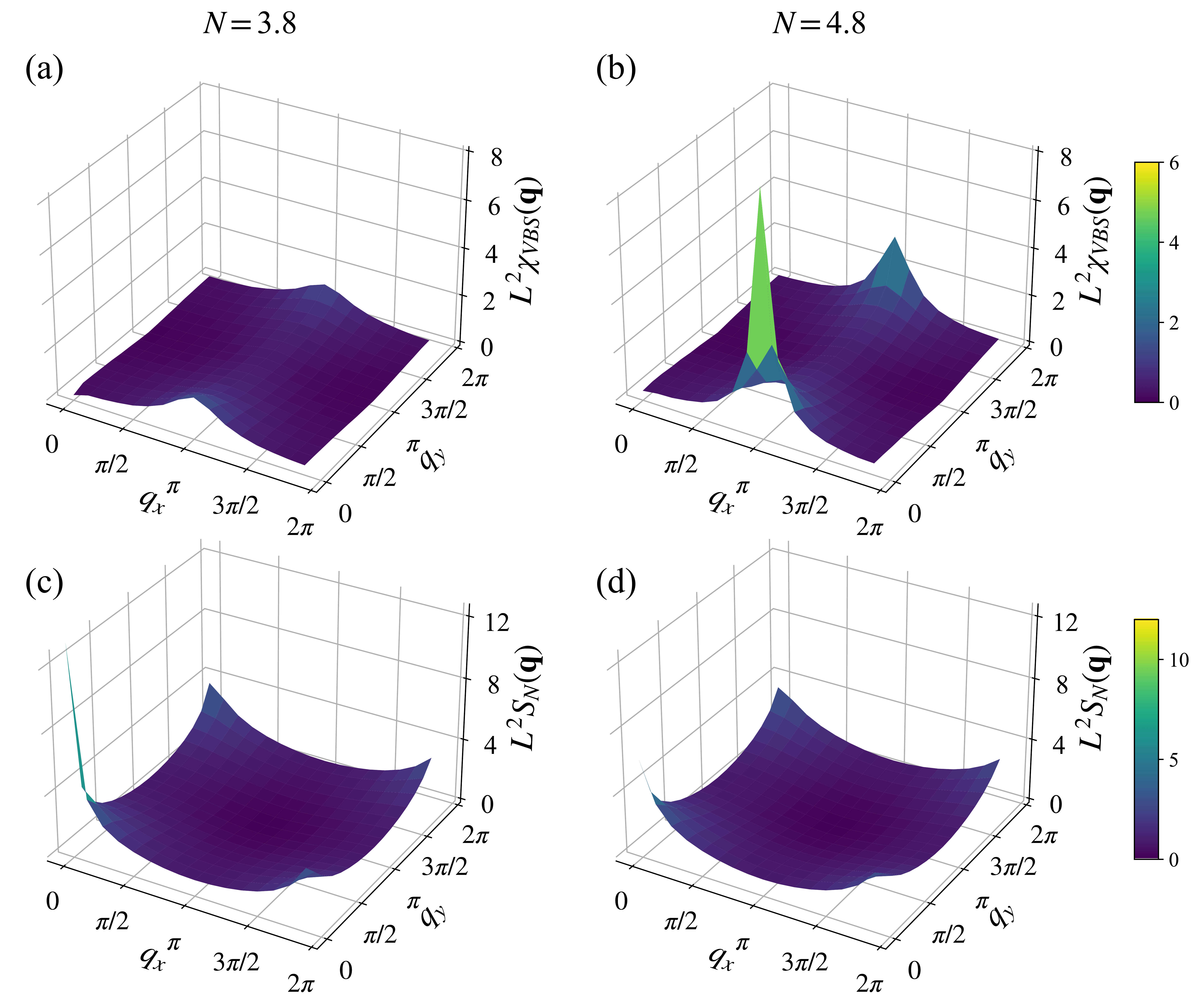}
\caption{\textbf{QMC data at linear system size $L=16$ for VBS susceptibility 
 $\chi_{V}(\mathbf{q})$ and spin structure factor $S_N(\mathbf{q})$ at different phases}.}
\label{fig:strcture_factor}
\end{figure}

\textcolor{black}{As an example, Fig.~\ref{fig:strcture_factor} shows QMC data at linear system size $L=16$ for the VBS susceptibility $\chi_V$ and spin structure factor $S_N$ inside the first Brillouin zone at two typical $N$ values. It is clear that $\chi_V$ develops a peak at $\mathbf{q}=(\pi,0)$ at $N=4.8$, which is inside the VBS phase, in contrast to the weak signal at $N=3.8$, the N\'eel phase. Similarly, $S_N$ develops a peak at $\mathbf{q}=(0,0)$ at $N=3.8$ which vanishes at $N=4.8$.}

\section{Incremental approach of EE computation}
\label{sec:appc}
\begin{figure}[htp!]
\includegraphics[width=0.8\columnwidth]{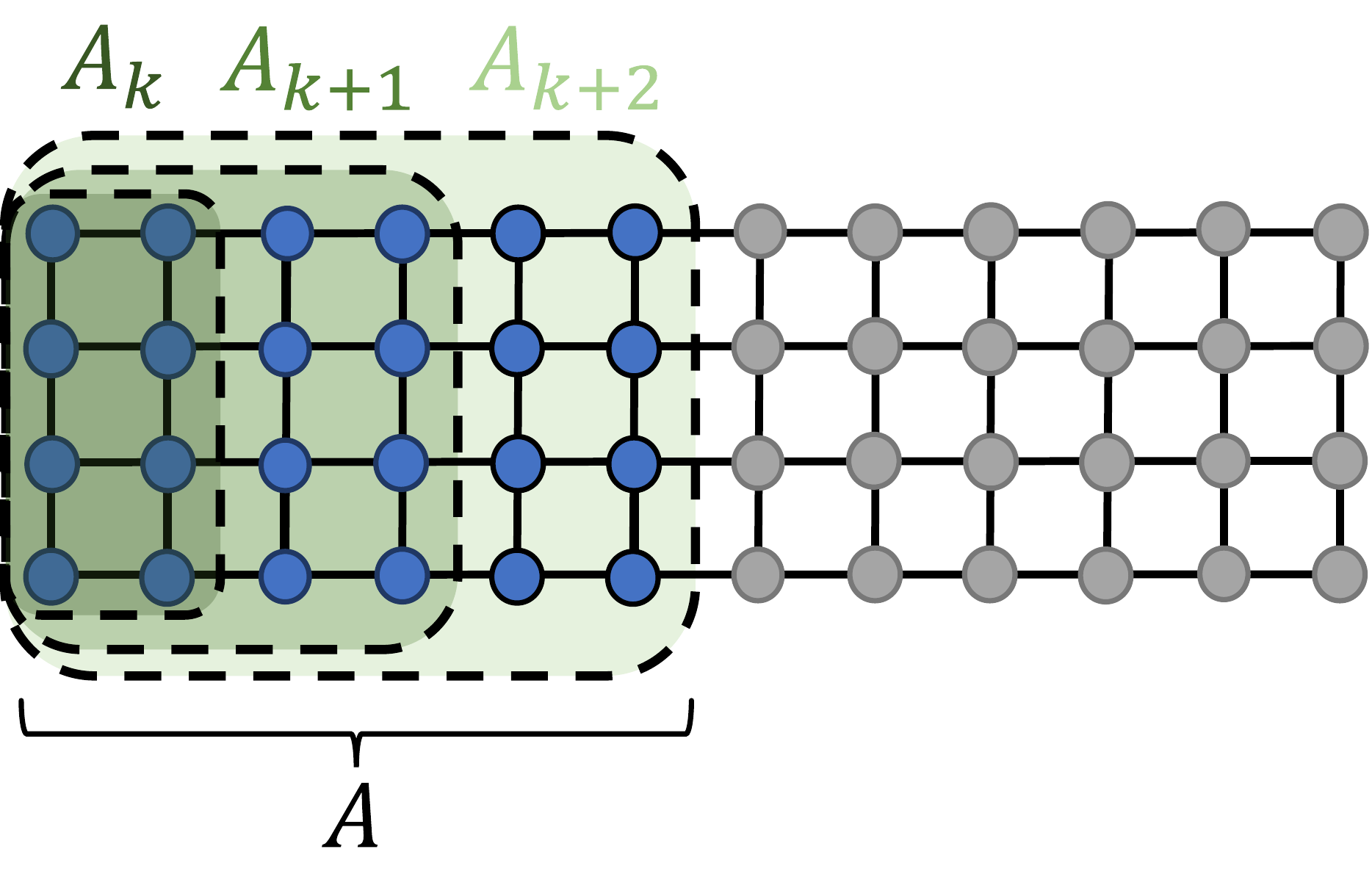}
\caption{\textbf{Schematic of the incremental entanglement regions.} $A_k$, $A_{k+1}$ and $A_{k+2}$ are the intermediate entanglement regions in which the ratio in Eq.~\eqref{eq:S1} are computed independently. Since the incremental regions are not very different from each other, the incremental ratios computed are well-conditioned, compared with direct computation of $\langle P_{\emptyset\rightarrow A}\rangle$ and $\langle P_{A\rightarrow \emptyset}\rangle$.}
\label{fig:icmt}
\end{figure}

Instead of calculating the probabilities $\langle P_{\emptyset\rightarrow A}\rangle$ and $\langle P_{A\rightarrow \emptyset}\rangle$ directly, the R\'enyi entanglement entropy $e^{-S_A^{(2)}}$ can also be broken down into smaller pieces, that is,
\begin{equation}
\begin{split}
e^{-S_A^{(2)}}&=\frac{{Z}_A}{{Z}_\emptyset }\\
&=\frac{Z_{A_{n}}}{Z_{A_{n-1}}}\frac{{Z_{A_{n-1}}}}{Z_{A_{n-2}}}\cdots\frac{Z_{A_k}}{{Z_{A_{k-1}}}}\cdots\frac{{Z_{A_1}}}{{Z_{A_0}}} \\
&=\frac{P_{A_{n-1}\rightarrow A_{n}}}{P_{A_{n}\rightarrow A_{n-1}}}\cdots\frac{P_{A_{k-1}\rightarrow A_{k}}}{P_{A_{k}\rightarrow A_{k-1}}}\cdots\frac{P_{A_{0}\rightarrow A_{1}}}{P_{A_{1}\rightarrow A_{0}}}
\end{split}
\label{eq:S1}
\end{equation}
where $A_n$ is set to be $A$, and $A_0$ is set to be $\emptyset$. $A_k$ are intermediate entanglement regions as shown in Fig.~\ref{fig:icmt}. They are added so that each probability has a larger value, and the statistical error can be suppressed.

In this study, we let each consecutive $A_k$ be two columns wider than the former $A_{k-1}$.

\section{Robustness of the anomalous log-correction near the quantum critical point}
\label{sec:appD}
\textcolor{black}{In this section, we analyze the robustness of the anomalous log-correction $b$ near the $N_c$ by (1) tracking the value of $b$ with different smallest system sizes attained $L_{\mathrm{min}}$ in fitting to bypassing the finite-size effect, and (2) scaling the EE at different $N$ values within the error bar of $N_c$ to demonstrate the anomalous $b$ value at $N=4.43$ is not a coincidence.}

\textcolor{black}{Table~\ref{table:tabS1} summarizes the fitted $b$ with different $L_{\mathrm{min}}$ choices. It is clearly shown that the fitted $b$ exhibits no drastic changes while increasing $L_{\mathrm{min}}$, despite the enlargement of the error bar due to fewer data points being included in the fitting. More importantly, all those $b$ values remain positive, implying the existence of the anomalous log-correction at the thermodynamic limit.}

\begin{table}[thb]
\begin{tabular}{c|c|l|c}
\hline
\hline

                & $L_{\rm{min}}=8$ & $L_{\rm{min}}=12$                  & $L_{\rm{min}}=16$ \\ \hline
\makecell{Fitted $b$ at \\ $N=4.43$}& 1.76(4)& \multicolumn{1}{c|}{1.6(2)} & 1.6(7)\\ \hline
\hline
\end{tabular}
\caption{\textbf{Fitted log-correction $b$ in at $N=4.43$ with a smooth cut.} The second, third, and fourth columns present the fitted $b$ with $L_{\rm min}=8, 12, 16$, respectively.}
\label{table:tabS1}
\end{table}

\begin{figure}[htp!]
\includegraphics[width=\columnwidth]{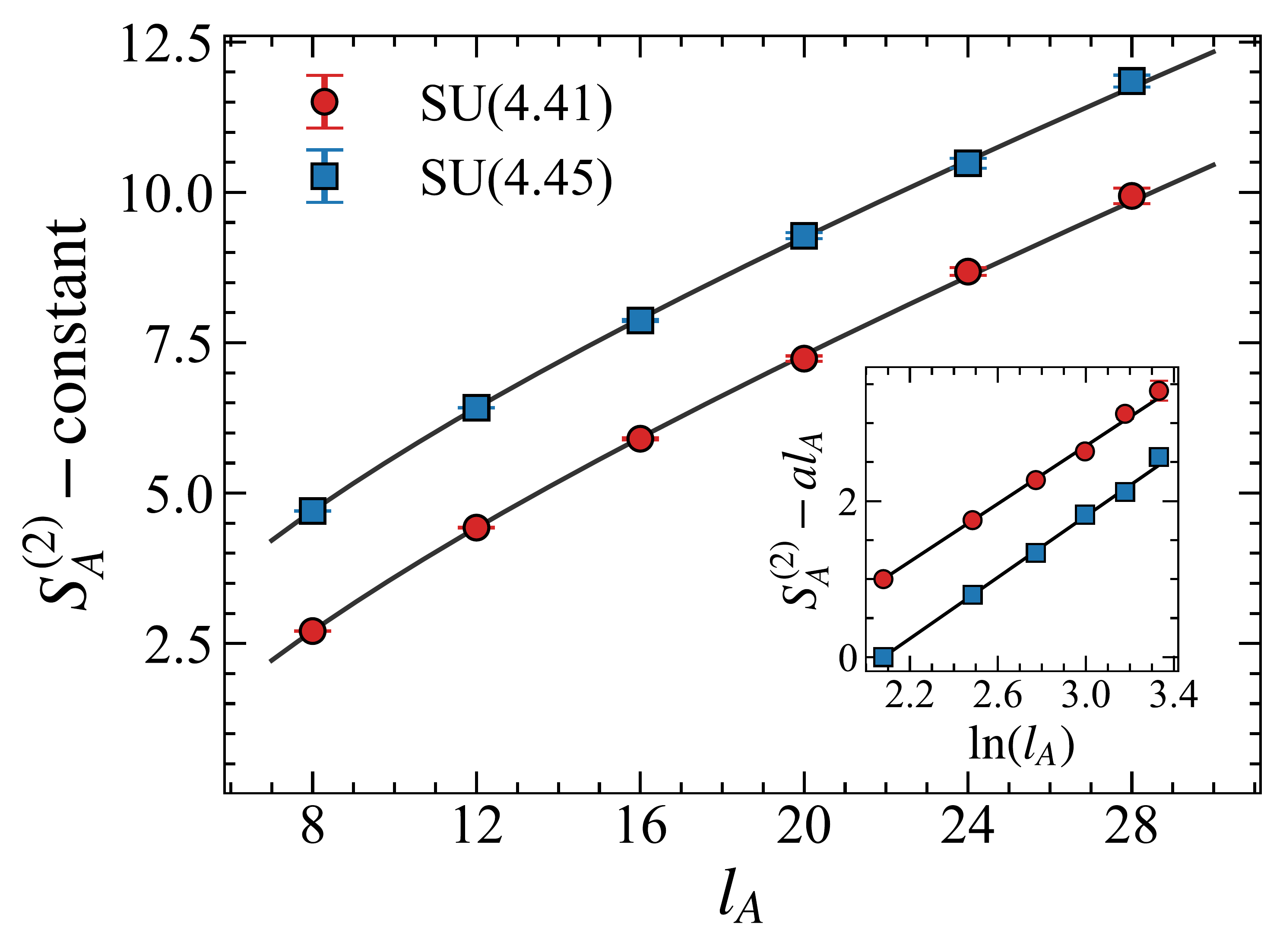}
\caption{\textbf{Scaling of entanglement entropy within a narrow window around $N_c$}. The fitted log-correction at SU(4.41) and SU(4.45) are $b=1.8(1)$ and $b=1.9(1)$ respectively. The linear behavior of $S^{(2)}-al_A$ against $\ln l_A$, as depicted in the inset, also confirms the existence of a finite log-correction.}
\label{fig:EE_supp}
\end{figure}

\textcolor{black}{Fig.~\ref{fig:EE_supp} provides the scaling of EE at two $N$ values within the error bar of the determined $N_c$. At both $N=4.41$ and $N=4.45$, the anomalous log-correction clearly manifests. The magnitude of the log corrections is also close to that at $N=4.43$, shown in Fig.~\ref{fig:fig5}. Therefore, the anomalous log-correction occurs in the entire interval of possible values of $N_c$, justifying that the positive $b$ at $N=4.43$ is not a coincidence. }

\newpage
\bibliographystyle{longapsrev4-2}
\bibliography{bibtex}

\end{document}